 \documentclass[12pt,preprint]{aastex}

\shorttitle{No QSO Redshift Periodicities or Association with
Galaxies}
\shortauthors{TANG $\&$ ZHANG}

\begin{document}


\title{Critical Examinations of QSO Redshift Periodicities and Associations with Galaxies in Sloan Digital Sky Survey Data}


\author{Su Min Tang\altaffilmark{1} and Shuang Nan Zhang\altaffilmark{1,2,3,4}}
\email{tangsm99@mails.tsinghua.edu.cn; zhangsn@tsinghua.edu.cn}

\altaffiltext{1}{Department of Physics and Center for
Astrophysics, Tsinghua University, Beijing 100084, China}
\altaffiltext{2}{Department of Physics, University of Alabama in
Huntsville, Optics Building 201C, Huntsville, AL 35899}
\altaffiltext{3}{Space Science Laboratory, NASA Marshall Space
Flight Center, SD50, Huntsville, AL 35812}
\altaffiltext{4}{Institute of High Energy Physics, Chinese Academy
of Sciences, P.O. Box 918-3, Beijing 100039, China}

\begin{abstract}
We have used the publicly available data from the Sloan Digital
Sky Survey and 2dF QSO Redshift Survey to test the hypothesis that
QSOs are ejected from active galaxies with periodic
non-cosmological redshifts. For two different intrinsic redshift
models, namely the Karlsson $\log(1+z)$ model and Bell's
decreasing intrinsic redshift (DIR) model, we do two tests
respectively. First, using different criteria, we generate four
sets of QSO-galaxy pairs and find there is no evidence for a
periodicity at the predicted frequency in $\log(1+z)$, or at any
other frequency. We then check the relationship between high
redshift QSOs and nearby active galaxies, and we find the
distribution of projected distance between high redshift QSOs and
nearby active galaxies and the distribution of redshifts of those
active galaxies are consistent with a distribution of simulated
random pairs, completely different from Bell's previous
conclusion. We also analyze the periodicity in redshifts of QSOs,
and no periodicity is found in high completeness samples, contrary
to the DIR model. These results support that QSOs are not ejected
from active galaxies.
\end{abstract}


\section{Introduction}

The debate on whether QSOs are ejected from the nuclei of
low-redshift galaxies with a periodic non-cosmological
``intrinsic" redshift has been going on for many years. Some
evidence has been claimed to suggest such an intrinsic redshift
hypothesis, in which QSOs have redshifts that are much larger than
their parent galaxies and the excess of redshift is assumed to
represent an always redshifted intrinsic component \citep{bb67,
arp90, kar90, chu98, bn01, bell04b, arp05}.

Two models have been discussed in the literature which predict
exact values for the preferred redshifts. One of these is the
Karlsson formula which suggests a periodicity of $\triangle
\log(1+z_{eff})=0.089$ with peaks lying at 0.061, 0.30, 0.60,
0.96, 1.14, 1.96 and so on (Karlsson 1977, 1990; Arp et al. 1990,
2005; Burbidge $\&$ Napier 2001, 2003), where $z_{eff}$ is the
redshift of the QSO measured relative to the nearby galaxy, called
effective redshift, which is defined as:
\begin{equation}
1+z_{eff}=(1+z_Q)/(1+z_G)
\end{equation}
where $z_Q$ is the observed quasar redshift and $z_G$ is the
redshift of the associated galaxy which is assumed to be the
ejecting galaxy. To explain such a periodicity, they claimed that
quasars are ejected by active galaxies and the putative parent
galaxies are generally much brighter than their quasar off-springs
\citep{arp05}. As claimed by Burbidge $\&$ Napier (2001, 2003),
the typical projected association separation is about 200 kpc.

Another model, namely decreasing intrinsic redshift model (DIR
model), was proposed by Bell (2004), where the QSO intrinsic
redshift equation is given by the relation:
\begin{equation}
z_{iQ}=z_f(N-M_N)
\end{equation}
where $z_f=0.62\pm0.01$ is the intrinsic redshift constant, $N$ is
an integer, and $M_N$ varies with $N$ and is a function of a
second quantum number $n$. In the DIR model, galaxies are produced
continuously through the entire age of the universe, and QSOs are
assumed to be ejected from the nuclei of active galaxies and
represent the first very short lived stage ($10^7-10^8$ yr) in the
evolution of galaxies \citep{bell04b}.

The above an intrinsic redshift hypothesis, if true, will have
far-reaching consequences for cosmology and the nature of QSOs.
Most of those previous studies on the Karlsson formula used rather
small samples (except for Arp et al. 2005), and have been
suspected that the claimed peaks were due to artifacts associated
with selection effects \citep{basu05}. To avoid such a
heterogeneous selection manner as well as personal prejudice,
Hawkins et al. (2002) tested the periodicity in $\log(1+z_{qso})$
with 2dF redshift survey data with 67291 nearby galaxies and 10410
QSOs; it was found that there is no periodicity in
$\log(1+z_{qso})$. However, Napier $\&$ Burbidge (2003) argued
that in order to use the 2dF sample to properly test the original
hypothesis, it is necessary to establish for each pair that the
galaxy is at least a late-type active spiral system. Arp et al.
(2005) also re-examined the 2dF sample and claimed that they found
that the redshifts of brighter QSOs in the QSO density contours
fit very exactly the long standing Karlsson formula and confirm
the existence of preferred values in the distribution of quasar
redshifts.

In an attempt to resolve these issues, we turn to the Sloan
Digital Sky Survey (SDSS) (and also 2dF QSO Redshift Survey (2QZ)
occasionally) to carry out this study, which have the largest
homogeneous sample of data  as well as the spectroscopic
sub-classification of galaxies. In section 2, to test whether
there is a periodicity existing in $\log(1+z)$, we construct four
sets of QSO-galaxy pairs with different QSOs and galaxies, with
all QSOs projected within 200 kpc from nearby galaxies at these
galaxies' distances. QSO density contours are also presented to
show that there is no periodicity in SDSS QSOs under such
analysis. In section 3, we examine the relationship between high-z
QSOs and nearby active galaxies to show that such QSOs are not
likely to be ejected by nearby active galaxies. In section 4, we
analyze the redshift distribution of QSOs in SDSS DR3 and 2QZ to
show that there is no evidence for non-artificial periodicity in
redshifts of QSOs, contrary to the DIR model. Discussion and
conclusion are described in Section 5.

\section{No Periodicity in $\log(1+z)$}

\subsection{The SDSS Data and Pair Selection}
In this section, we use the SDSS DR1 QSO catalog \citep{sch03} and
the New York University Value-Added Galaxy Catalog (NYU-VAGC)
\citep{bla04}. For reliability in the derived redshifts, we
consider only those QSOs in the range of $z>0.4$, and galaxies in
the range of $0.01<z<0.2$ with the highest plate quality labeled
as ``good" and with no redshift warning. This quality control
leaves a total of 190591 galaxies and 15747 QSOs in the sample.

Three issues need to be carefully addressed when we analyze the
relation between foreground galaxies and QSOs, as well as the
redshift distribution of QSOs, since due to the survey strategies
and the instrumental limitations, the selections of galaxies and
QSOs are not entirely independent, and the selection of QSOs in
SDSS is also dependent on redshift. First, due to the mechanical
constraint in SDSS that spectroscopic fibers must be separated by
$55''$ on a given plate (Blanton et al. 2003), consequently some
QSO-galaxy pairs would be missing from the spectroscopic sample.
However, this issue would have little effect on the results for
two reasons: (1) $55''$ corresponds to an angular distance of $40$
kpc for a galaxy at $z=0.04$, which is the typical value of
redshift in our galaxy sample, only few pairs (about $5\%$) would
be missing in a given separation of 200 kpc for randomly
distributed QSOs and galaxies, which is also shown in Fig.~7 where
the distribution of data pairs are consistent with randomly
distributed pairs; (2) such fiber constraint is independent of
redshifts of galaxies or QSOs, therefore its redshift distribution
will not be biased, although some pairs are missing in the sample.
Second, the magnitude limits of the SDSS galaxy and QSO
spectroscopic surveys are quite different, i.e., $i<19.1$ for
$z<3$ QSOs, $i<20.2$ for $z>3$ QSOs, $r<17.77$ for most sampled
galaxies and $r<19.5$ for luminous red galaxies (Richards et al.
2002; Strauss et al. 2002), hence the magnitudes of QSOs are
mostly higher than galaxies. However, since in the ejection
hypothesis, the parent galaxies are generally much brighter than
their QSO off-springs (Bell 2004b; Arp et al. 2005), the pair
making process are not likely to be affected by the magnitude
differences between QSOs and galaxies, which is also shown in
Fig.~12 and Fig.~13. Moreover, the fact that the completeness of
the spectroscopic selection varies with redshift (Richards et al.
2002) will consequentially affect the redshift distribution of
QSOs and might cause artificial periodicities into data, as will
be discussed extensively in Section 4. However, since low-z
($z<2.5$) QSOs which have flat and high completeness level
($>90\%$; Richards et al. 2002) occupied a very large fraction of
all QSOs (about $90\%$), such selection dependence in redshift
would not smear out intrinsic periodicities in QSO redshift if
they do exist.

It has been suggested that quasars with bright apparent magnitude
will generally be nearby and the redshifts of these quasars will
require little or no correction for the periodicity effects to be
manifested, and those low redshift galaxies with which such
quasars appear to be preferentially associated, tend to be
morphologically disturbed active galaxies \citep{arp05}.
Therefore, to make our results more compelling, we select a
sub-sample of 3724 QSOs, called bright QSOs, which have i-band
magnitudes less than 18.5, and a sub-sample of 77426 galaxies,
called active galaxies, which are labeled as starforming,
starburst, starforming broadline or starburst broadline galaxies
in the spectroscopic sub-classification. Then we construct four
sets of QSO-galaxy samples by intercrossing them, and get four
sets of pairs in which a QSO is projected within 200 kpc from a
galaxy: 4572 pairs for QSO-nearby galaxies, 3216 pairs for
QSO-active nearby galaxies, 1129 pairs for bright QSO-nearby
galaxies and 791 pairs for bright QSO-active nearby galaxies. When
there is more than one galaxy within the 200 kpc projected
distance limit of the QSO, we take the closest galaxy in projected
distance to make up the pair.

\subsection{Analysis and Results}

We make power spectrum analysis to investigate the periodicity
hypothesis  of Karlsson (1977). The power $I$ is defined as in
Burbidge $\&$ Napier (2001):
\begin{equation}
I(\nu)=2R^2/N,
\end{equation}
where
\begin{equation}
R^2=S^2+C^2,
\end{equation}
with
\begin{equation}
S=\sum^{N}_{i=1} w_i \sin(2\pi x_i/P),  C=\sum^{N}_{i=1} w_i
\cos(2\pi x_i/P),
\end{equation}
with $\nu=1/P$ and $x_i=\log(1+z_i)$. Here $w_i$ is a weighting
function, and $w_i\equiv 1$ except in section 4, as in the
analysis of Burbidge $\&$ Napier (2001). For randomly and
uniformly distributed redshifts, $\overline {I}=2$.

To test our code developed for this study, we first re-analyze the
290 QSOs in Karlsson and Napier $\&$ Burbidge's data sets
\citep{kar90,bn01,bn03}, as shown in  Fig.~1. Errors on $I(P)$ are
given by using the bootstrap methods \citep{efr79} in the
following steps: (1) we take the non-zero number in each bin in
the upper histogram as the expectation value of a Poisson
distribution; (2) we re-sample each bin following the Poisson
distribution to re-produce 1000 new sets of data, repeat the power
spectral analysis on these re-samplings, and finally calculate the
standard deviations in the derived values of $I$ at different
periods $P$. Clearly the periodicity at around $\triangle
\log(1+z_{eff})=0.089$ is highly significant at above 3.5$\sigma$
level.

In Figs.~2-5 we show histograms of the effective redshifts of QSOs
paired with galaxies and their unwindowed power spectra with
standard deviations calculated in the same way as for Fig.~1.
Pairs in these four figures are for QSO-nearby galaxies,
QSO-active nearby galaxies, bright QSO-nearby galaxies and bright
QSO-active nearby galaxies, as described in Sec. 2.1. Our results
show that for these significantly larger samples than that in
Fig.~1, all peaks appeared in the power spectra are consistent
with Poisssonian fluctuations, i.e., there is no evidence for a
periodicity at the predicted frequency in $\log(1+z)$, or at any
other frequency.

\subsection{QSO Density Contours}
After the work of Hawkins et al. (2002) on 2dF data, Arp et al.
(2005) argued that the predicted periodic redshifts are apparent
in the brighter 2dF quasars in the QSO density contours. We
therefore use SDSS DR1 data to construct the contours defined by
Arp et al. (2005), where the whole region is divided into boxes
$\bigtriangleup z \times \bigtriangleup B =0.075 \times 0.3$ in
the redshift/apparent magnitude plane, then the number of quasars
in each box is counted. To show whether the predicted periodic
redshifts are obscured by our coarse grid sizes, a contour with
$\bigtriangleup z \times \bigtriangleup B =0.05 \times 0.2$ is
also presented for comparison. As shown in Fig.~6, the peak
positions are consistent in the two contour plots, and there is no
evidence for redshift peaks at the predicted positions.

\section{No Strong Connection between Active Galaxies and Bell's High-z QSOs}

In Bell (2004b), a high-redshift QSO sample from SDSS and a
low-redshift QSO and QSO-like object sample from Hewitt $\&$
Burbidge (1993) were presented. Though the dips at redshifts of
2.7 and 3.5 have been explained as being caused by the lower
efficiency of the selection algorithm at these redshifts
\citep{ric02}, Bell (2004b) nevertheless claimed that the
corresponding redshift peaks at 3.1 and 3.7 in the high-z SDSS
QSOs come from the intrinsic redshift broadening by Doppler
ejection and Hubble flow components, which is in favor of the DIR
model. Through analysis of the profiles of such peaks, Bell
(2004b) derived a mean cosmological components to be $z_c \sim
0.066$ for the high-z sample.

In the DIR model, galaxies are produced continuously through the
entire age of the universe, and QSOs are assumed to be ejected
from the nuclei of active galaxies and represent the first very
short lived stage ($10^7-10^8$ yr) in the evolution of galaxies.
If this is true, there must be some connection between foreground
active galaxies and high-z QSOs beyond gravitational lensing. Here
we examine the high-redshift samples taken from the SDSS data. We
test the relationship between 2691 QSOs with redshifts in
2.4$\sim$4.8 and 77426 nearby active galaxies with redshifts in
0.01$\sim$0.2 from NYU-VAGC, all of which have the highest plate
quality labeled as ``good" and with no redshift warning. We
inter-compare these two data sets to find all QSO-galaxy pairs
within an angular separation corresponding to less than a given
distance $D_{separation}$ from several kpc to 1 Mpc at the
distance of the galaxy. In some cases, there is more than one
galaxy within the $D_{separation}$ projected distance limit of the
QSO; for these objects we take the closest galaxy in projected
distance to make up the pair. Since it is suggested that all QSOs
are born out of active galaxies and QSOs should be significantly
fainter than their parent galaxies (Bell 2004b), we would not miss
a considerable fraction of parent active galaxies for high-z QSOs
if the DIR model is right.

The distribution of projected separation distance for all pairs is
shown in Fig.~7, and the redshift distribution of active galaxies
in pairs with QSOs is shown in Fig.~8. Both of them are consistent
with random distributions, but totally different from the
distribution from the ejection simulation with a ejection velocity
of 11,000 km s$^{-1}$ which was given by Bell (2004b) as typical
values, and the mean redshift of these galaxies is $z\sim0.044$,
also significantly different from Bell's result of 0.066. Here the
random distributions are obtained by moving the positions of all
galaxies by 1 degree in random directions; thus these galaxies
should be completely unrelated to background QSOs. The ejection
simulation is done by ejecting all QSOs from randomly selected
active galaxies with three ejection velocities: 11,000 km
s$^{-1}$, 40,000 km s$^{-1}$ and 80,000 km s$^{-1}$, and with a
uniformly distributed age of $0 \sim 10^8$ yr which is given by
Bell (2004b) as a typical value.

To quantitatively show the differences between simulations and
`true' QSO-galaxy pairs, i.e., pairs found in the data but not
necessarily physical pairs, results of chi-squared tests are given
in Table 1. In $3\sigma$ confidence level for both distributions
of projected separation distance and redshift distribution of
active galaxies in pairs with QSOs, the `true' QSO-galaxy pairs
are consistent with random distributions, but inconsistent with
ejection hypothesis with ejection velocity up to 80,000 km
s$^{-1}$.

\begin{table}
\begin{center}
\caption{Results of chi-squared tests for the distribution of
projected separation distance and the redshift distribution of
active galaxies in pairs with high-z QSOs between `true' pairs and
simulations. \label{tbl}}

\begin{tabular}{c|ccc|ccc}
\tableline\tableline

 simulations & $\chi^2$ \tablenotemark{a}& $\chi^2$/N \tablenotemark{a}& p \tablenotemark{a} & $\chi^2$ \tablenotemark{b}& $\chi^2$/N \tablenotemark{b}& p \tablenotemark{b}\\
\tableline
randomly distributed galaxies & 38.456 & 1.6720 & 0.023 & 18.818 & 0.9904 & 0.47 \\
ejected QSOs, v=11,000 km/s & 2560.3 & 111.3174 &  $<10^{-10}$  & 994.10 & 52.3211 &  $<10^{-10}$  \\
ejected QSOs, v=40,000 km/s & 227.12 & 9.8748 &  $<10^{-10}$  & 265.31 & 13.9637 &  $<10^{-10}$  \\
ejected QSOs, v=80,000 km/s & 85.496 & 3.7172 & $4.0\times10^{-9}$ & 144.37 & 7.5984 &  $<10^{-10}$  \\
\tableline
\end{tabular}

%% Any table notes must follow the \end{tabular} command.

\tablenotetext{a}{Column 2-3 are for the distribution of projected
separation distance. Pairs with projected separation distance less
than 60 kpc (first and second points in Fig.~7) are not used in
the tests to avoid the SDSS 55'' fiber constraint, and all
simulations are normalized to get a minimum $\chi ^2$. The degrees
of freedom are $N=23$.}

\tablenotetext{b}{ Column 4-7 are for the redshift distribution of
active galaxies in pairs with high-z QSOs. All the first zero
point in Fig.~8 is not used in the tests and all simulations are
normalized to get a minimum $\chi ^2$. The degrees of freedom is
$N=19$.}

\end{center}
\end{table}

\section{No Periodicity in $z$}

We also analyze the periodicity in redshifts of QSOs in SDSS DR3
(Schneider et al. 2005) and 2dF (Croom et al. 2004) to investigate
in larger database of QSOs whether there is a periodicity of
$\Delta z = 0.67 \pm 0.05$, predicted by the DIR model. Six data
sets are used in this section: all 46,420 QSOs in SDSS DR3
(Fig.~9), 22,497 QSOs with the highest quality flag in 2dF
(Fig.~10), a high completeness (close to $100\%$) sub-sample
containing 23,109 QSOs with Galactic-extinction corrected $i$-band
magnitude ($m_i$) less than 19 and redshift less than 2 in SDSS
DR3 (Richards et al. 2002) (Fig.~11(a)), and three sub-samples
containing QSOs in low completeness (less than $50\%$) regions in
SDSS DR3: 15,696 QSOs with $m_i>19$ and $z<2.4$ (Fig.~11(b)),
19,064 QSOs with $m_i>19$ in all redshifts (Fig.~11(c)), and 9,763
QSOs with $z>2$ (Fig.~11(d)). To reduce the edge effect produced
by the truncated redshift distribution which has a lower redshifts
cut-off due to the small space volume sampled and a higher
redshifts cut-off due to the observational flux limit (see e.g.
Hawkins et al. 2002), we follow Hawkins et al. to use the Hann
function as a weighting in equation 5,

\begin{equation}
w_i=\frac{1}{2}[1-\cos(\frac{2\pi x_i}{L})],
\end{equation}
where $L$ is chosen to cover the range of redshifts. Here $L=5.4$
for the full SDSS sample, $L=3.1$ for the 2dF sample, and $L=1.95,
2.1, 5.1$ and 3.4 for the four SDSS sub-samples respectively.

After smoothing off the sharp edges in the lowest and highest
redshifts, a periodicity around $\Delta z = 0.67$ is detected in
the full sample of SDSS QSOs, as shown in Fig.~9; however a
periodicity of $\Delta z = 0.67 \pm 0.05$ or any other frequency
is not found in the 2dF QSOs, as shown in Fig.~10. Such a
difference between these two surveys is not surprising since the
redshift-dependent spectroscopic completeness is relatively flat
in 2dF (Croom et al. 2004), while in SDSS the spectroscopic
completeness varies drastically at some redshifts (Richards et al.
2002). It is therefore improper to use all QSO redshifts in SDSS
to probe any intrinsic periodicity without addressing selection
bias. To further investigate whether such a periodicity around
$\Delta z = 0.67$ in SDSS QSOs is spuriously produced by various
incompleteness as function of redshift, we select a
high-completeness sub-sample of 23,109 QSOs with $m_i<19$ and
$z<2$ in SDSS DR3, and three sub-samples containing QSOs in
low-completeness regions. As shown in Fig.~11, no periodicity is
found in the high-completeness sample where the power spectrum is
consistent with a continuously ascending curve due to the low
frequency component of the redshift distribution, whereas in
different low-completeness samples, strong periodicity always
appears, but with different peak locations (0.88 in (b), 0.67 in
(c) and 0.74 in (d)). This should be a strong indicator that the
peaks in low-completeness samples are caused by different
selection effects in different samples. In sum, there is no
evidence for intrinsic periodicity in redshifts of QSOs.

\section{Discussion and Conclusion}

However, one might ask whether it is because we have some paired QSOs with wrong parent
galaxies so that not only the effective redshifts of QSOs show no periodicity, but also
high-z QSOs and nearby active galaxies show no strong connection. The wrong-pairing
indeed could happen that when there is more than one galaxy within the $D_{separation}$
projected distance limit of the QSO and we take the closest galaxy in projected
distance to make up the pair. In the following we quantitatively examine this
possibility and its effect.

For the pair making process in section 2, the $D_{separation}$ is
200 kpc, which is less than the average projection distance
between QSOs and galaxies, and the percentage of cases in which
there are two or more galaxies within the projected distance is
27$\%$ for QSO-nearby galaxies, 19$\%$ for QSO-active nearby
galaxies, 27$\%$ for bright QSO-nearby galaxies and 18$\%$ for
bright QSO-active nearby galaxies, respectively. This means that
for a majority of paired QSOs ($>73 \%$), there is only one galaxy
within the given projected distance and would not be paired
incorrectly, hence the claimed periodicity should have been
detected in our larger samples if they did exist.

For the pair making process in section 3, the largest $D_{separation}$ is 1 Mpc which
is larger than the average projection distance between QSOs and galaxies ($\sim$400
kpc), therefore wrong-pairing may occur more frequently here. It will be even worse if
the typical ejection distance is larger than the mean projection separation of QSOs and
active galaxies. So would such wrong pairs result in the good agreement between ejected
model and randomly generated pairs? We answer this question by making the following
test. Suppose that all QSOs are ejected by randomly selected active galaxies with a
 given ejection velocity (11,000 km s$^{-1}$, 40,000 km s$^{-1}$ and 80,000 km
 s$^{-1}$), and with a uniformly distributed age of $0 \sim
10^8$ yr which is given by Bell (2004) as typical values, we get
200 sets of false pairs. As shown in Fig.~7-8 and Table 1, the
distribution of such simulated ejection QSO-galaxy pairs are
totally different from random distribution. We therefore conclude
that the random-like distribution of QSO-active galaxy pairs could
not be produced by the ejection model.

Another question is that whether we miss periodicities of QSOs by
setting a lower limit of $z=0.01$ for galaxies and no constraint
in QSO-galaxy magnitude relation. Though the lower limit of
$z=0.01$, the same as in Hawkins et al. (2002), is chosen to have
confidence in the derived angular distance, as well as avoid large
projection effect of very nearby galaxies, we re-analyze SDSS DR1
QSOs and galaxies again with no redshift limits on galaxies and
set magnitude constraint that all paired QSOs should be at least 5
or 3 magnitudes fainter than the paired galaxy. As shown in
Fig.~11 and Fig.~12, similar to our results in section 2.2, there
is no evidence for a periodicity at the predicted frequency in
log$(1+z)$, or at any other frequency.

In summary, using samples from SDSS and 2QZ, we demonstrate that
not only there is no periodicity at the predicted frequency in
$\log(1+z)$ and $z$, or at any other frequency, but also there is
no strong connection between foreground active galaxies and high
redshift QSOs. These results are against the hypothesis that QSOs
are ejected from active galaxies or have periodic intrinsic
non-cosmological redshifts.

\acknowledgments

We thank the anonymous referee and Dr. Bell for valuable
suggestions that have significantly improved this paper. This
study is supported in part by the Special Funds for Major State
Basic Research Projects and by the National Natural Science
Foundation and the Ministry of Education of China. SNZ also
acknowledges NASA for partial financial support through several
research grants.


%% copy editing.  Individual instruments can be provided in parentheses,
%% after the keyword, but they will not be verified.

% Facilities: \facility{Nickel}, \facility{HST(STIS)}, \facility{CXO(ASIS)}.

\clearpage

\begin{figure}
\epsscale{.80} \plotone{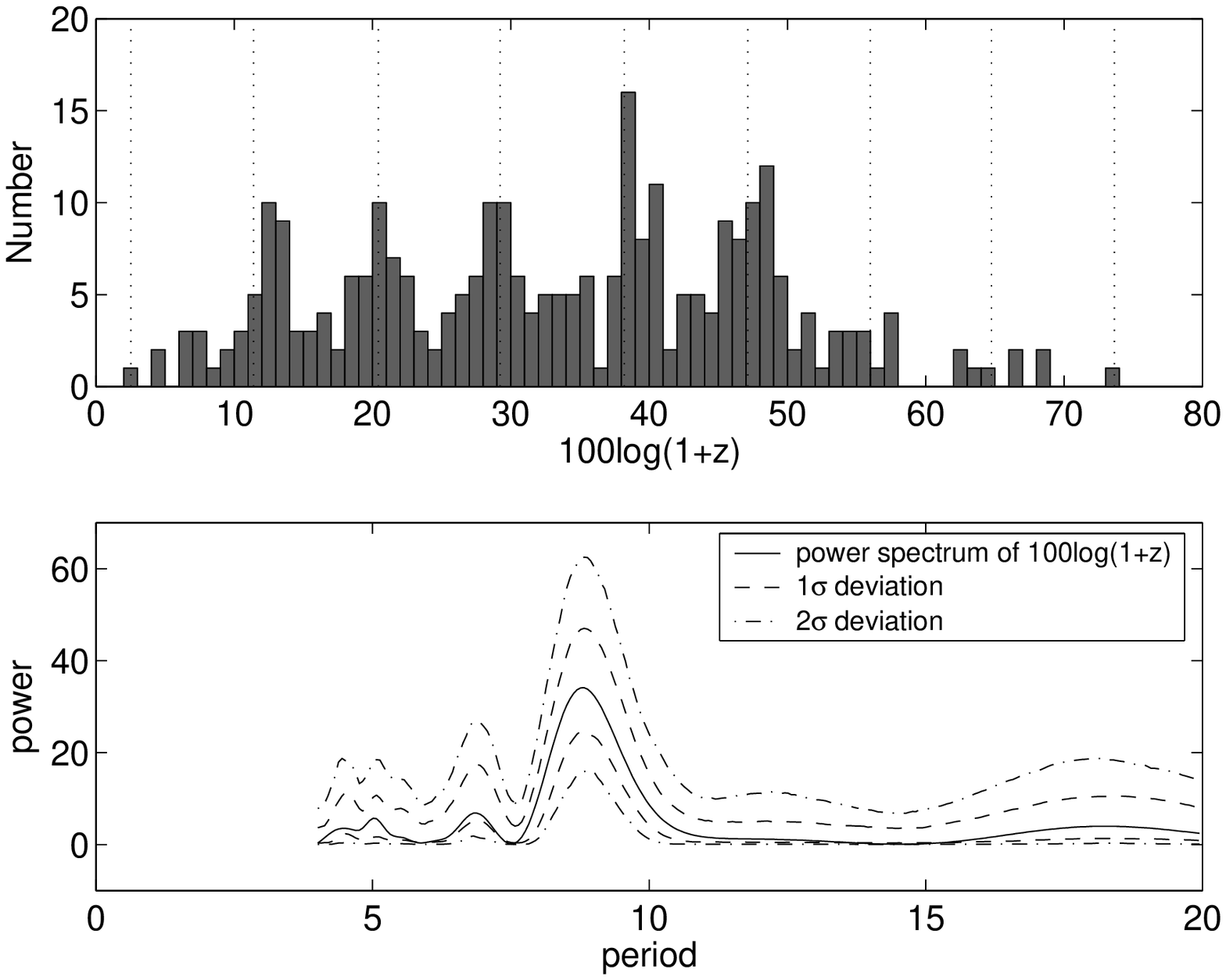} \caption{Combined QSO data
set from Karlsson (1990) and Burbidge $\&$ Napier (2001,
  2003). Upper panel: histogram of redshifts of these QSOs. Peaks predicted by Karlsson's
  formula are indicated by dotted vertical lines. Lower panel:
  unwindowed power spectra of $100\log(1+z)$ (solid line) with 1$\sigma$ (dash lines)
  and 2$\sigma$ (dash-dotted lines) deviations given from 1000
  bootstrap simulations.\label{fig1}}
\end{figure}
\clearpage

\begin{figure}
\epsscale{.80} \plotone{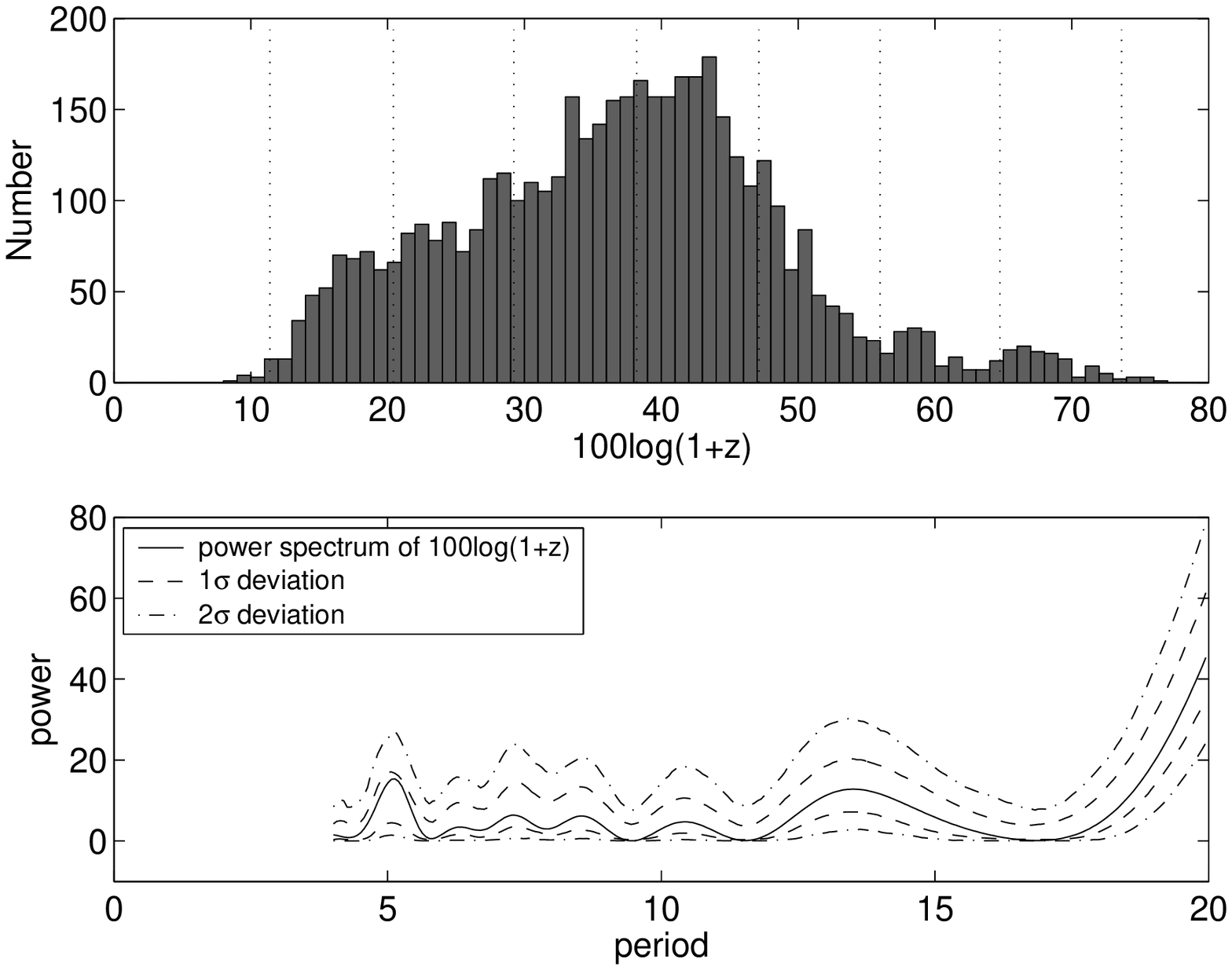} \caption{Effective redshifts of
4572 QSOs paired with nearby galaxies in a projection
  distance less than 200 kpc. Upper panel: histogram of redshifts of these QSOs. Peaks predicted by Karlsson's
  formula are indicated by dotted vertical lines. Lower panel:
  unwindowed power spectra of $100\log(1+z_{eff})$ (solid line) with 1$\sigma$ (dash lines)
  and 2$\sigma$ (dash-dotted lines) deviations given from 1000
  bootstrap simulations.\label{fig2}}
\end{figure}
\clearpage

\begin{figure}
\epsscale{.80} \plotone{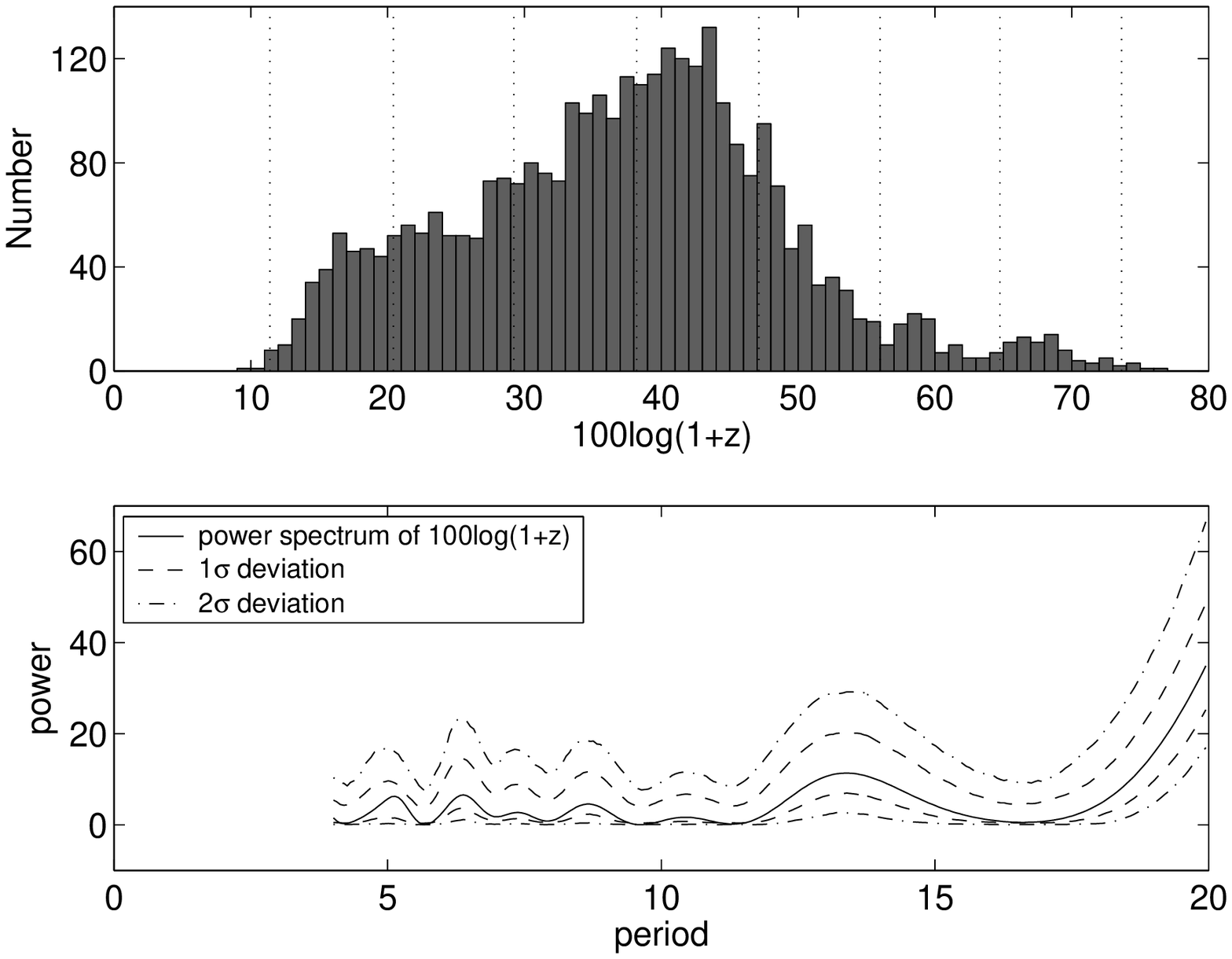} \caption{Effective redshifts of
3216 QSOs paired with nearby active galaxies (starforming or
starburst galaxies) in a projection
  distance less than 200 kpc. Upper panel: histogram of redshifts of these QSOs. Peaks predicted by Karlsson's
  formula are indicated by dotted vertical lines. Lower panel:
  unwindowed power spectra of $100\log(1+z_{eff})$ (solid line) with 1$\sigma$ (dash lines)
  and 2$\sigma$ (dash-dotted lines) deviations given from 1000
  bootstrap simulations.\label{fig3}}
\end{figure}
\clearpage

\begin{figure}
\epsscale{.80} \plotone{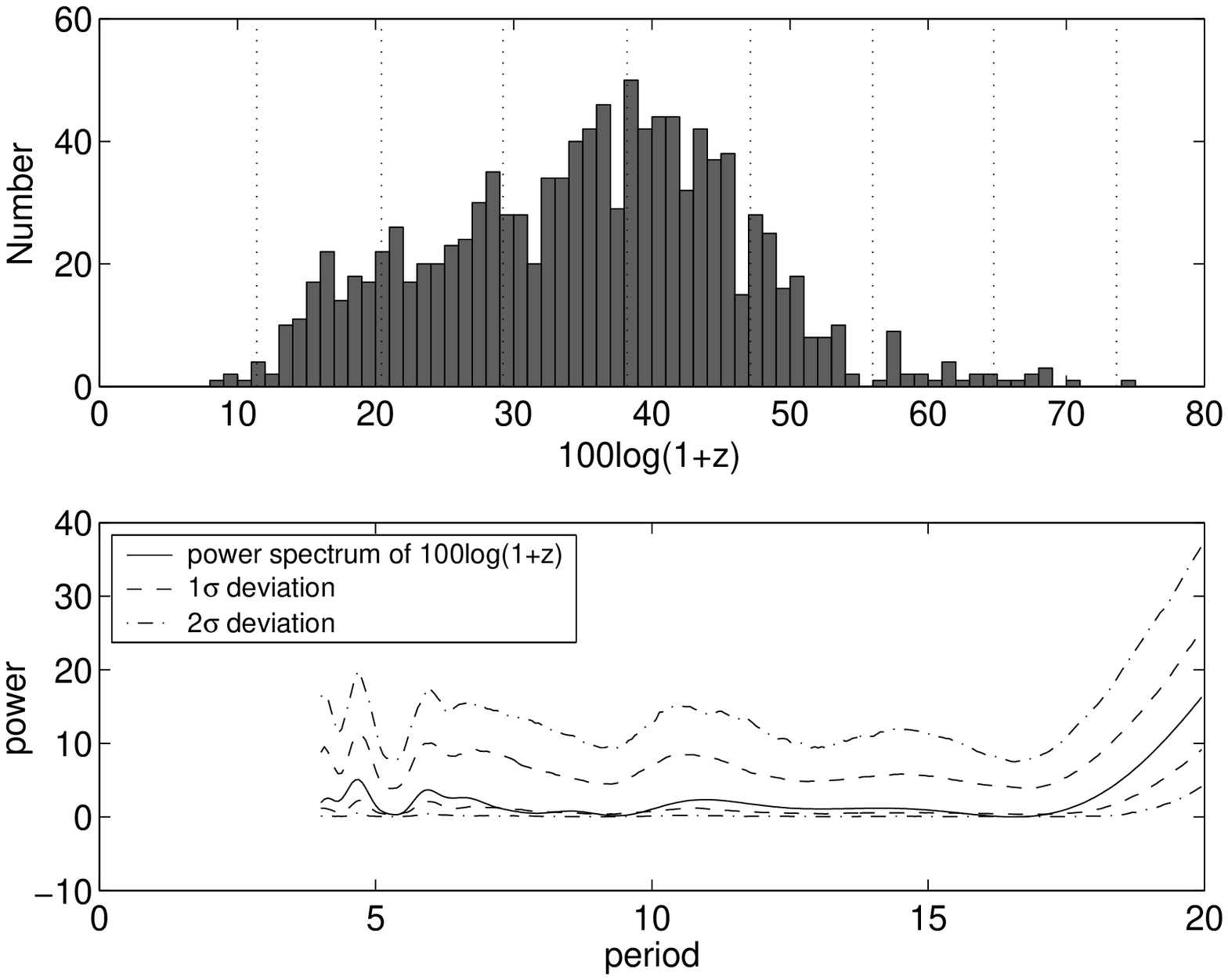} \caption{Effective redshifts of
958 bright QSOs ($i<18.5$) paired with nearby galaxies in a
projection
  distance less than 200 kpc. Upper panel: histogram of redshifts of these QSOs. Peaks predicted by Karlsson's
  formula are indicated by dotted vertical lines. Lower panel:
  unwindowed power spectra of $100\log(1+z_{eff})$ (solid line) with 11$\sigma$ (dash lines)
  and 2$\sigma$ (dash-dotted lines) deviations given from 1000
  bootstrap simulations.\label{fig4}}
\end{figure}
\clearpage

\begin{figure}
\epsscale{.80} \plotone{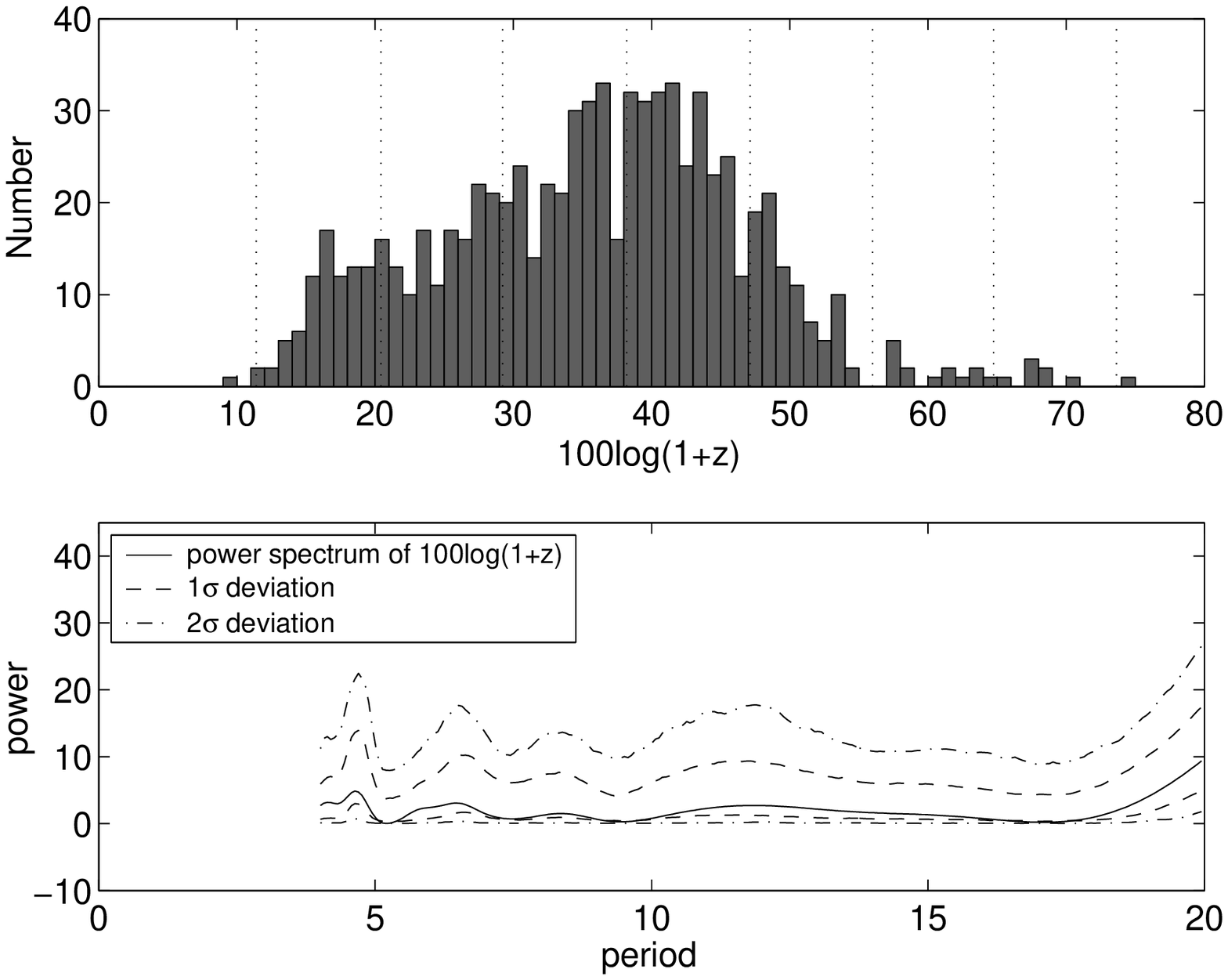} \caption{Effective redshifts of
671 bright QSOs ($i<18.5$) paired with nearby active galaxies
(starforming or starburst galaxies) in a projection
  distance less than 200 kpc. Upper panel: histogram of redshifts of these QSOs. Peaks predicted by Karlsson's
  formula are indicated by dotted vertical lines. Lower panel:
  unwindowed power spectra of $100\log(1+z_{eff})$ (solid line) with 1$\sigma$ (dash lines)
  and 2$\sigma$ (dash-dotted lines) deviations given from 1000
  bootstrap simulations.\label{fig5}}
\end{figure}
\clearpage

\begin{figure}
\plottwo{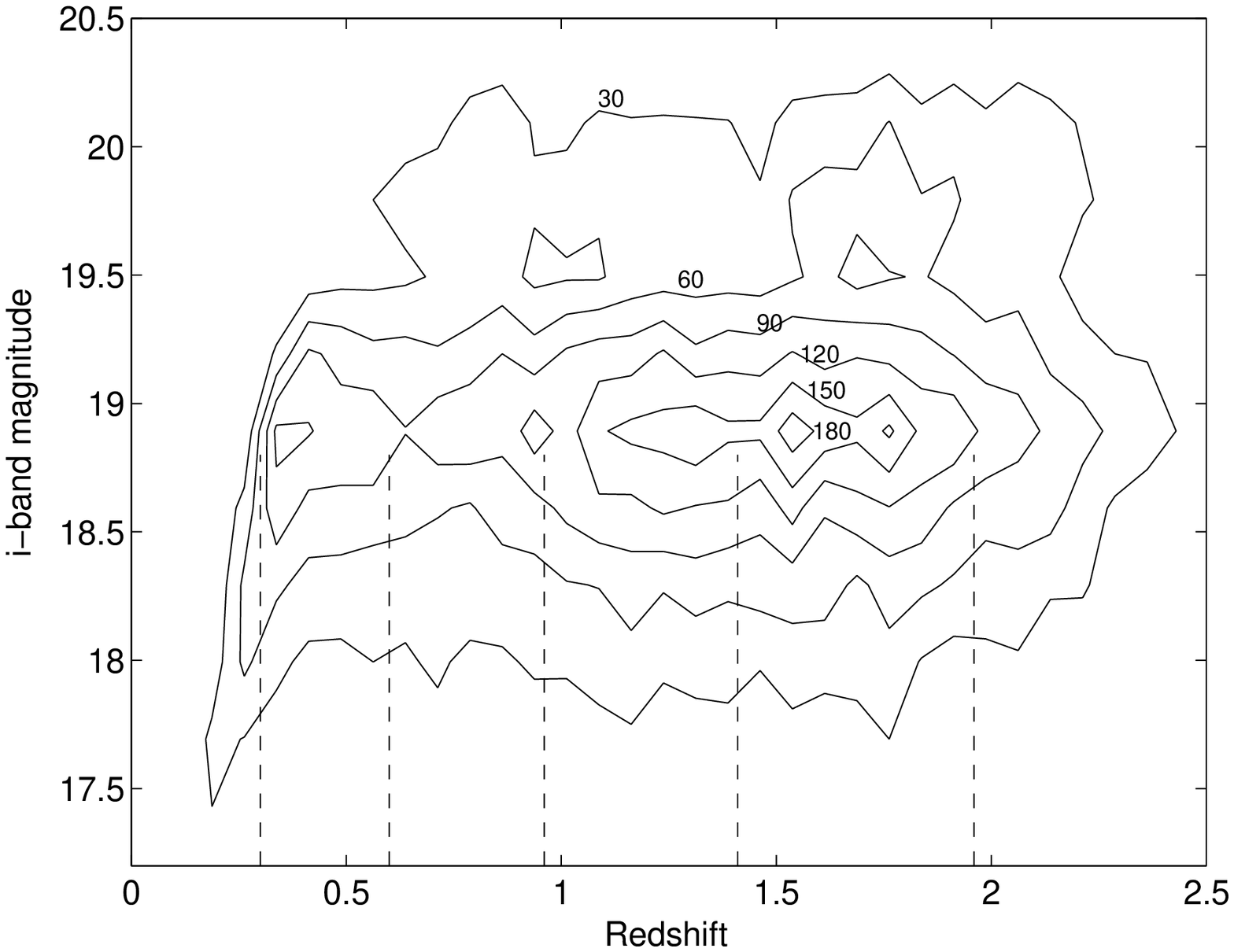}{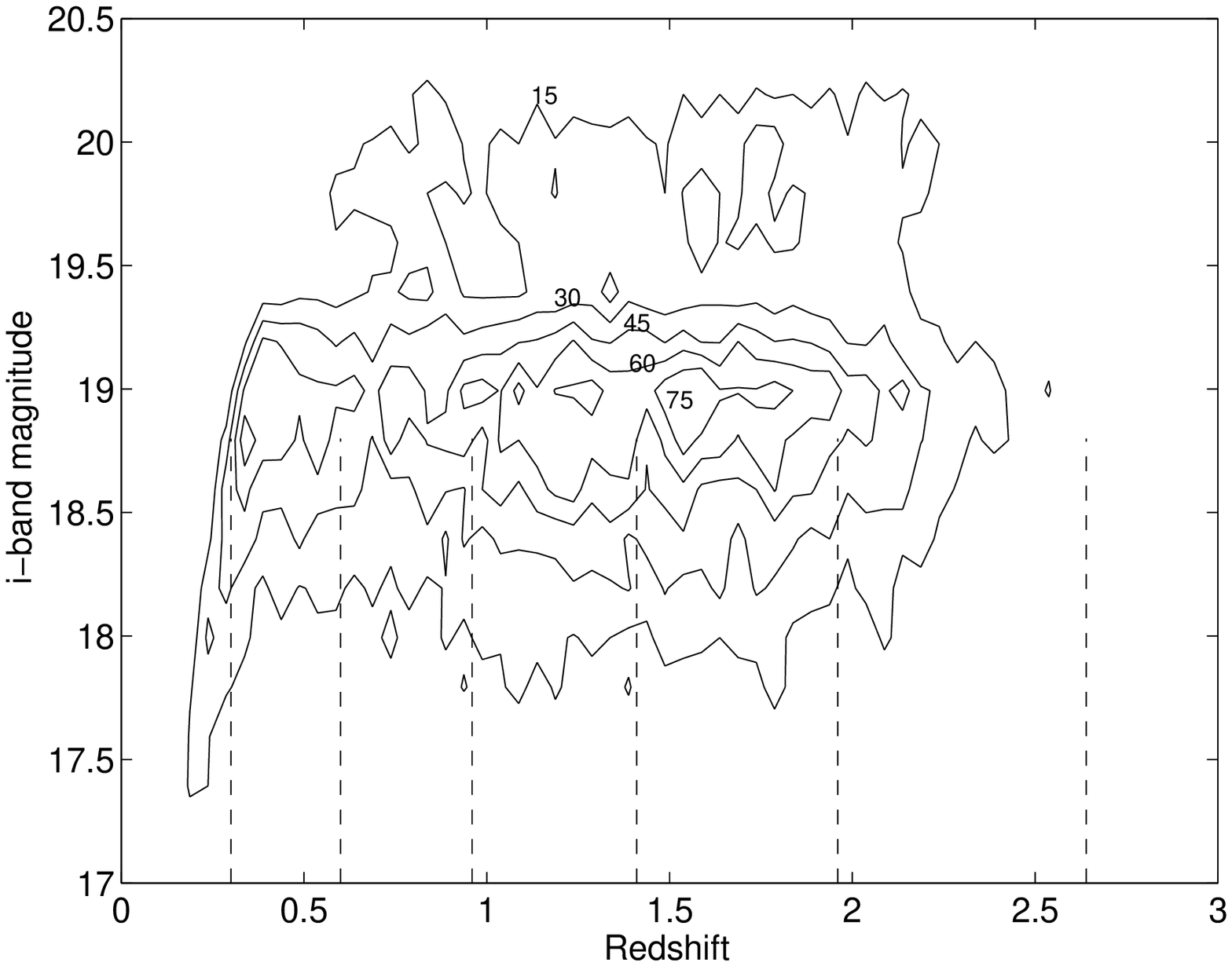} \caption{Apparent magnitude vs measured
redshift plot
  for QSOs in the SDSS DR1 catalog. In the left one, the whole region is divided into boxes $\bigtriangleup z \times
\bigtriangleup B =0.075 \times 0.3$ in the redshift/apparent
magnitude plane, while $\bigtriangleup z \times \bigtriangleup B
=0.05 \times 0.2$ in the right one. The contours represent
  QSO density in steps of 180, 150, 120, 90, 60 and 30 in the left one,
  while in the right one they represent 75, 60, 45, 30 and 15, from the
  innermost (high density) to outermost (low density). The predicted
  Karlsson peaks at z=0.30, 0.60, 0.96, 1.41, 1.96 and 2.64 are
  shown by vertical lines.\label{fig6}}
\end{figure}
\clearpage

\begin{figure}
\epsscale{.80} \plotone{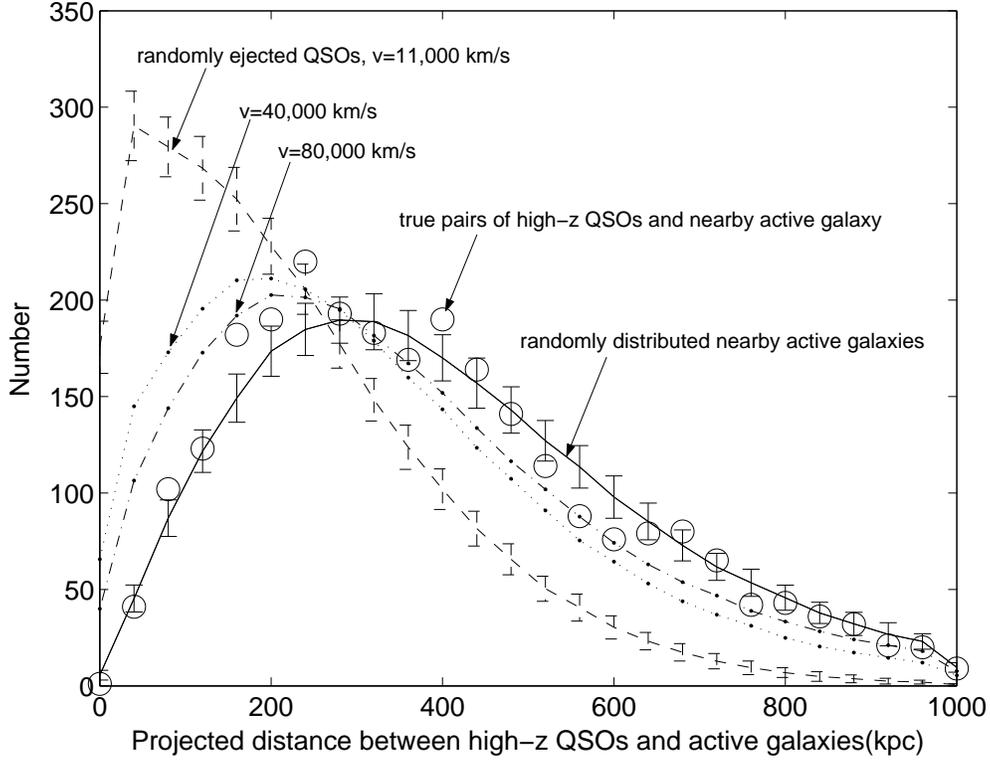} \caption{Distribution of
projected distance between
  2604 high-z QSOs ($2.4<z<4.8$) and their paired nearby active galaxies
  in NYU-VAGC. The circles represent `true' pairs, i.e., pairs found in the data, but not necessarily physical pairs.
   The solid line with error bars
  is the average of 200 simulations of QSOs and random distributed
  galaxies.
  Averages of 200 simulations of randomly ejected QSOs and active galaxies are also presented,
  where QSOs are produced by ejection from randomly chosen galaxies
  with a uniformly distributed age in $0-10^8$ yr and three different velocities:
  11,000 km/s for dash line with error bars, 40,000 km/s for dotted line with points and
  80,000 km/s for dash-dotted line with points. \label{fig7}}
\end{figure}
\clearpage

\begin{figure}
\epsscale{.80} \plotone{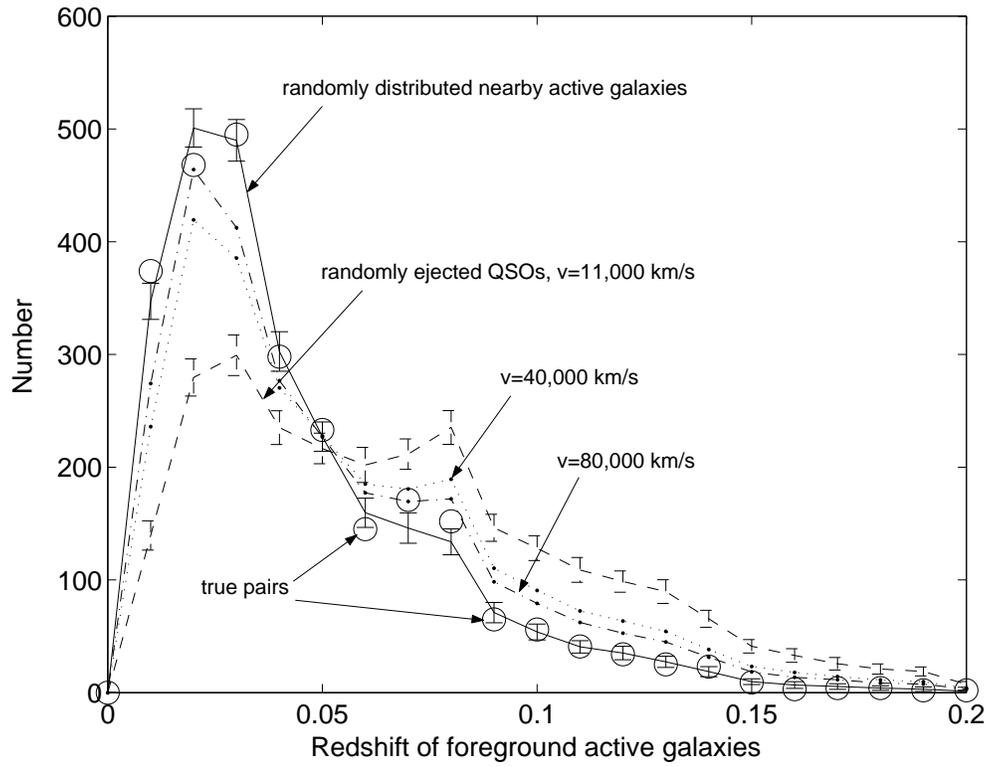} \caption{Distribution of
2604 foreground active galaxies which have
  at least one high-z QSO behind within a projected distance 1 Mpc in NYU-VAGC.
  The circles represent `true' pairs. Others
  are the same as in Fig.7.\label{fig8}}
\end{figure}
\clearpage

\begin{figure}
\epsscale{.80} \plotone{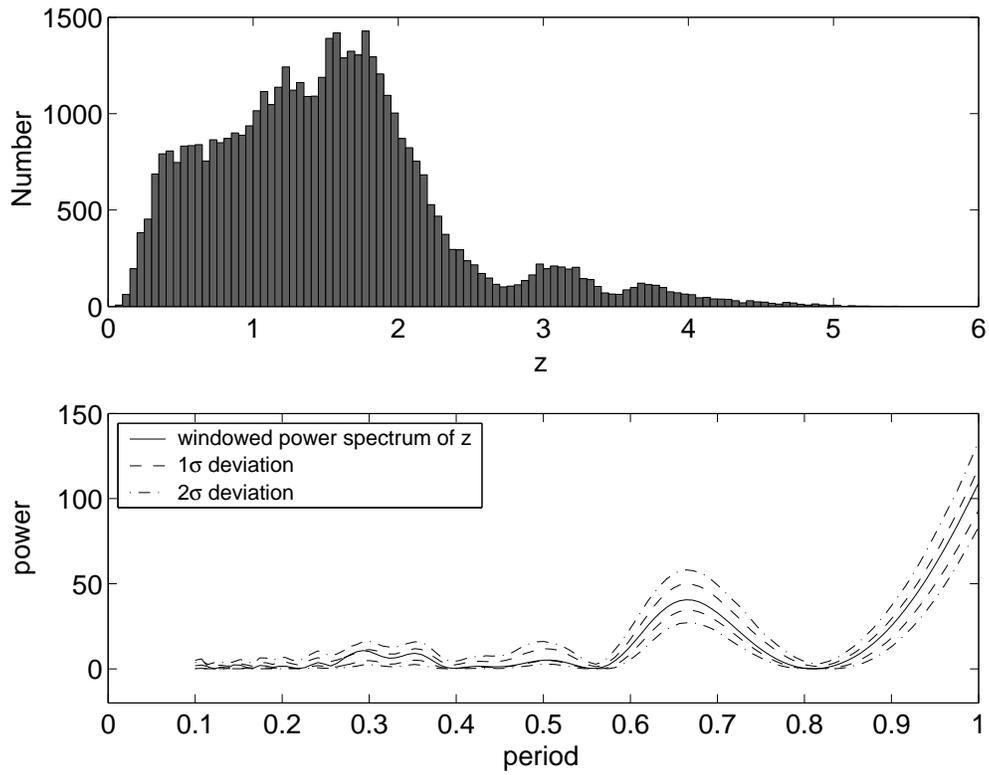} \caption{Redshifts of 46,420 QSOs
in SDSS DR3. Upper panel: histogram of redshifts of these QSOs.
Lower panel: power spectra of z (solid line) weighted using a Hann
function with 1$\sigma$ (dash lines) and 2$\sigma$ (dash-dotted
lines) deviations given from 1000 bootstrap simulations.
\label{fig9}}
\end{figure}
\clearpage

\begin{figure}
\epsscale{.80} \plotone{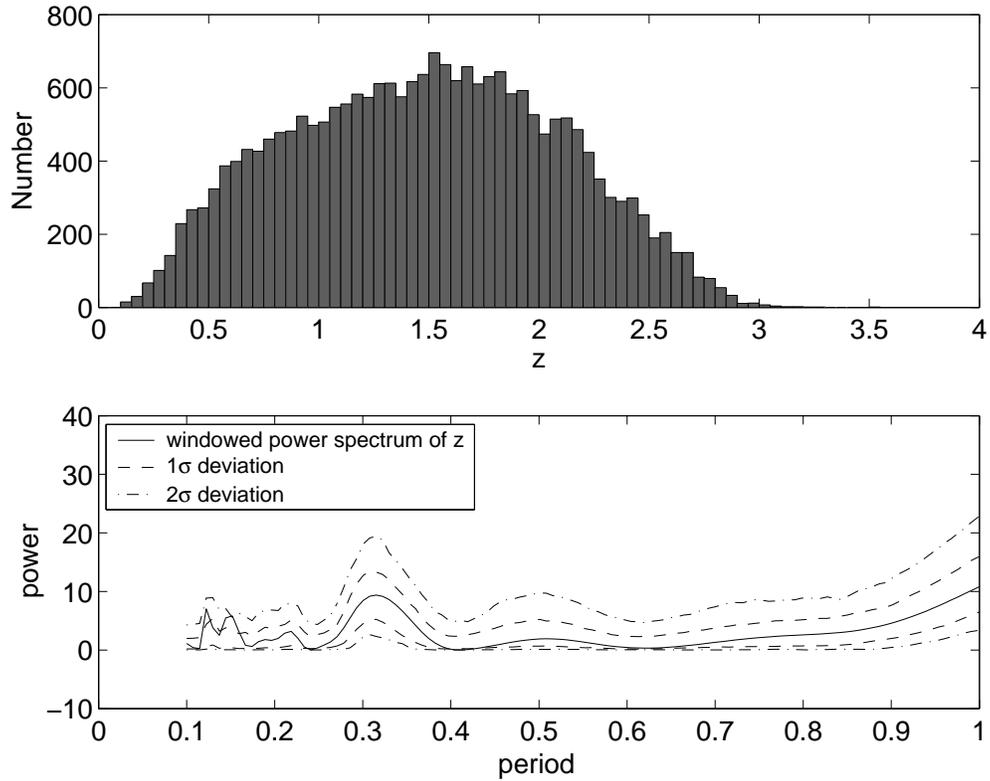} \caption{Redshifts of 22,497 QSOs
with the highest quality flag in 2dF. Upper panel: histogram of
redshifts of these QSOs. Lower panel: power spectra of z (solid
line) weighted using a Hann function with 1$\sigma$ (dash lines)
and 2$\sigma$ (dash-dotted lines) deviations given from 1000
bootstrap simulations. \label{fig10}}
\end{figure}
\clearpage

\begin{figure}
\epsscale{.80} \plotone{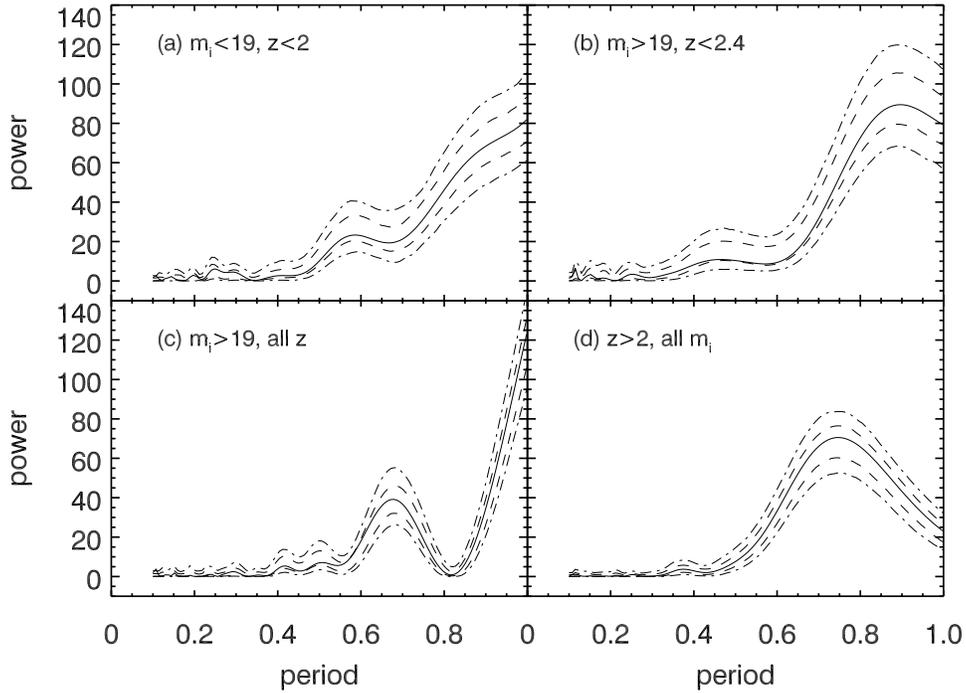} \caption{Power spectra of redshift
of four sub-samples from SDSS DR3. (a) is for the
high-completeness sample containing 23,109 QSOs with $m_i<19$ and
$z<2$, and others are for samples containing QSOs in
low-completeness regions: (b) is for 15,696 QSOs with $m_i>19$ and
$z<2.4$, (c) is for 19,064 QSOs with $m_i>19$, and (d) is for
9,763 QSOs with $z>2$. Power spectra of z (solid line) is weighted
using a Hann function with 1$\sigma$ (dash lines) and 2$\sigma$
(dash-dotted lines) deviations given from 1000 bootstrap
simulations. \label{fig11}}
\end{figure}
\clearpage

\begin{figure}
\epsscale{.80} \plotone{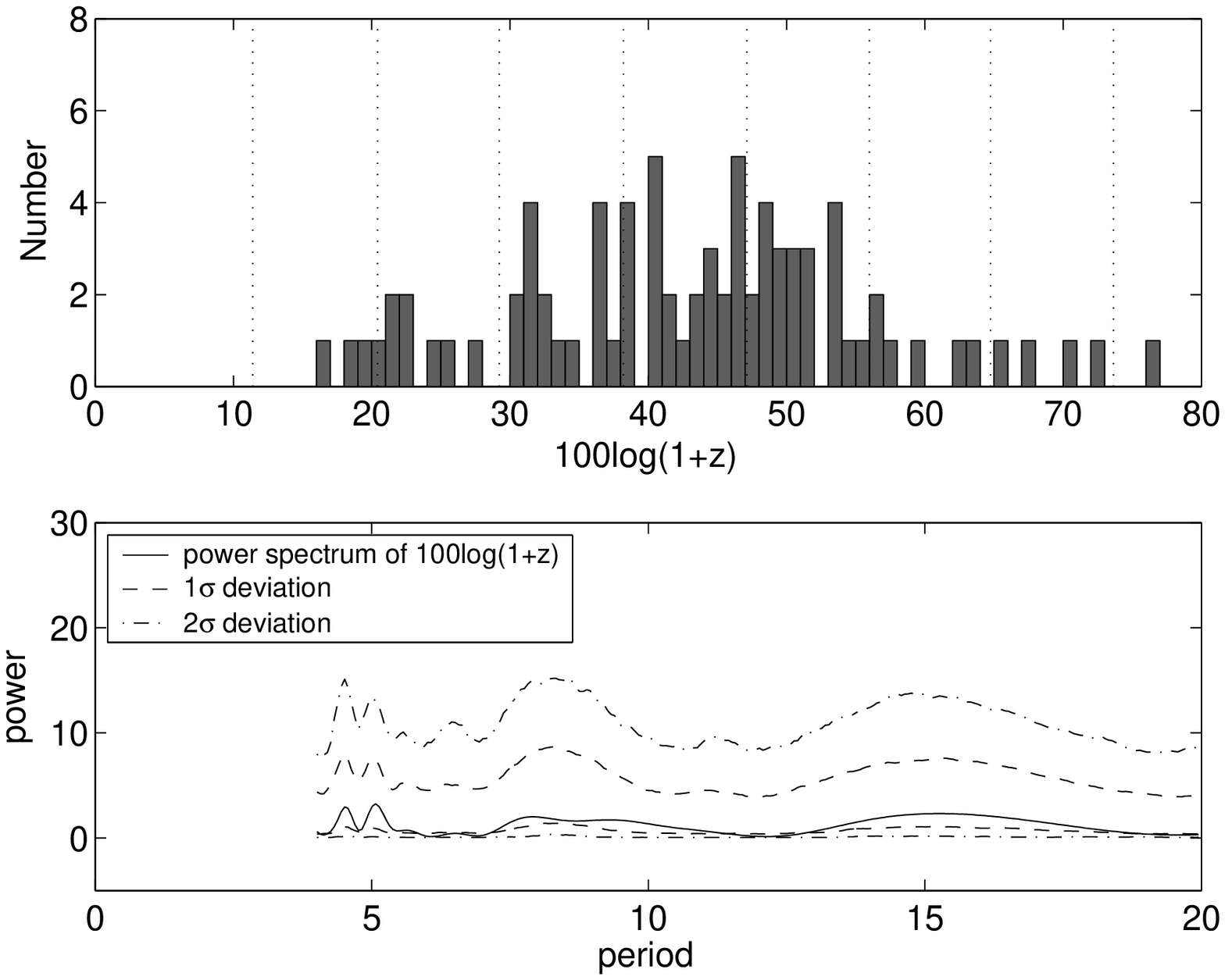} \caption{Effective redshifts of
82 QSOs paired with galaxies in a projection
  distance less than 200 kpc, which are at least 5 magnitudes fainter than the paired galaxies in $i$-band.
   Upper panel: histogram of redshifts of these QSOs. Peaks predicted by Karlsson's
  formula are indicated by dotted vertical lines. Lower panel:
  unwindowed power spectra of $100\log(1+z_{eff})$ (solid line) with 1$\sigma$ (dash lines)
  and 2$\sigma$ (dash-dotted lines) deviations given from 1000
  bootstrap simulations. \label{fig12}}
\end{figure}
\clearpage

\begin{figure}
\epsscale{.80} \plotone{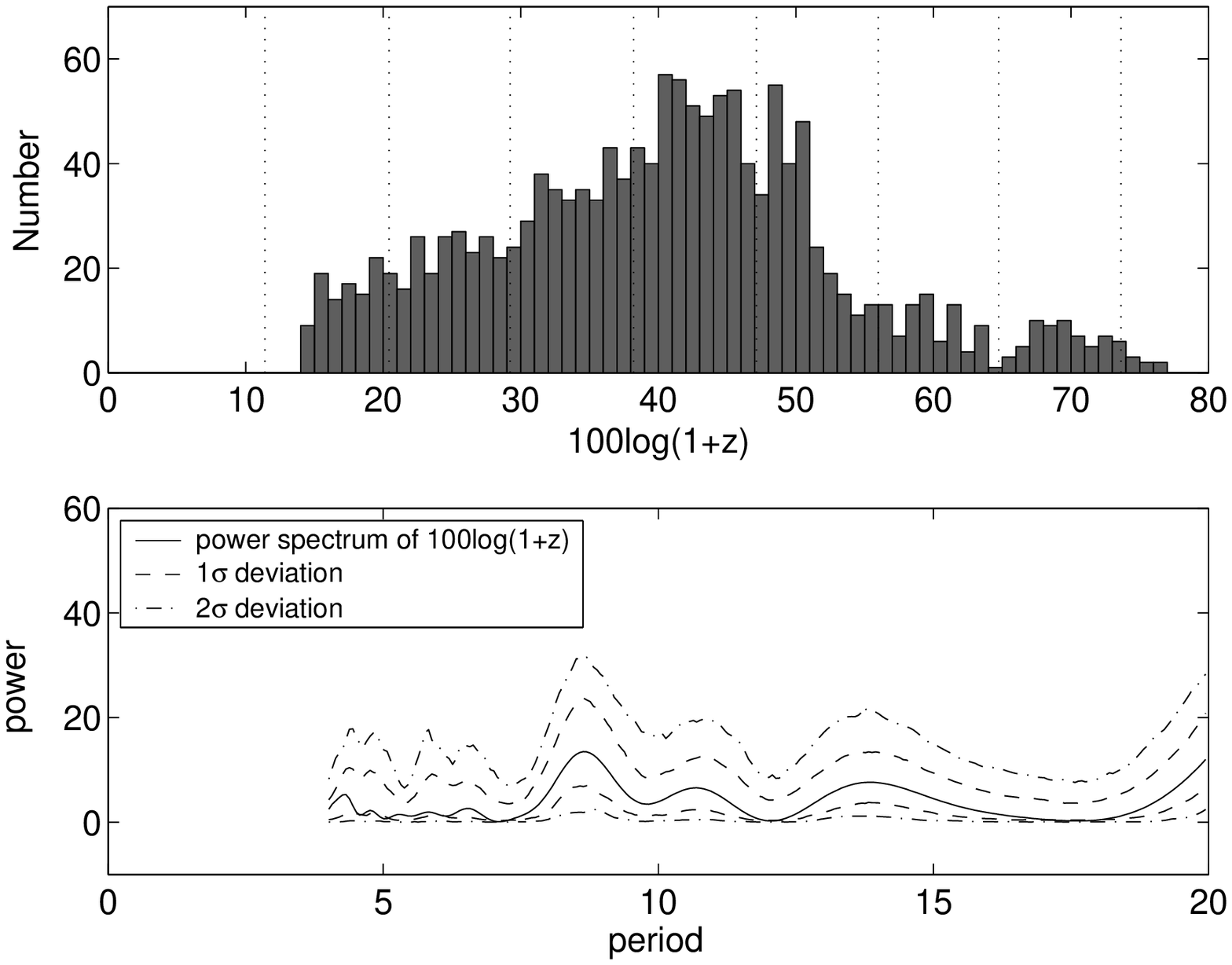} \caption{Effective redshifts of
1459 QSOs paired with galaxies in a projection
  distance less than 200 kpc, which are at least 3 magnitudes fainter than the paired galaxies in $i$-band.
  Upper panel: histogram of redshifts of these QSOs. Peaks predicted by Karlsson's
  formula are indicated by dotted vertical lines. Lower panel:
  unwindowed power spectra of $100\log(1+z_{eff})$ (solid line) with 1$\sigma$ (dash lines)
  and 2$\sigma$ (dash-dotted lines) deviations given from 1000
  bootstrap simulations.
\label{fig13}}
\end{figure}
\clearpage


\begin{thebibliography}{}
\bibitem[Abazajian et al.(2003)]{aba03}  Abazajian, k., et al. 2003, \aj, 126, 2081
\bibitem[Arp et al.(1990)]{arp90}  Arp, H., Bi, H. G., Chu, Y., \& Zhu, X.  1990,
\aap, 239, 33
\bibitem[Arp et al.(2005)]{arp05}  Arp, H., Roscoe, D., \& Fulton, C.  2005, astro-ph/0501090
\bibitem[Basu(2005)]{basu05} Basu, D.  2005, \apjl, 618, L71
\bibitem[Bell \& Comeau(2003)]{bell03} Bell, M. B. \& Comeau, S. P.
2003, submitted to ApJ (astro-ph/0305060)
\bibitem[Bell(2004a)]{bell04a} Bell, M. B. 2004a, submitted to ApJ (astro-ph/0403089)
\bibitem[Bell(2004b)]{bell04b} Bell, M. B.  2004b, \apj, 616, 738
\bibitem[Blanton et al.(2003)]{bla03} Blanton, M. R., et al. 2003,
AJ, 125, 2276
\bibitem[Blanton et al.(2005)]{bla04} Blanton, M. R., et al. 2005, accepted by AJ (astro-ph/0410166)
\bibitem[Burbidge \& Burbidge(1967)]{bb67} Burbidge, G. R. \& Burbidge, E. M.  1967, \apjl, 148,
L107
\bibitem[Burbidge \& Napier(2001)]{bn01} Burbidge, G. R. \& Napier, W. M.  2001, \apj, 121,
21
\bibitem[Burbidge(2003)]{bur03} Burbidge, G. R.  2003, \apj, 585,
112
\bibitem[Chu et al.(1998)]{chu98} Chu, Y., Wei, J., Hu, J., Zhu, X., \& Arp, H.  1998, \apj, 500,
596
\bibitem[Croom et al.(2004)]{cro04} Croom, S. M., et al. 2004,
\mnras, 349, 1397
\bibitem[Efron(1979)]{efr79} Efron, B.  1979, Ann. Stat., 7, 1
\bibitem[Gaztanaga(2003)]{gaz03} Gaztanaga, E. 2003, \apj, 589, 82
\bibitem[Hawkins et al.(2002)]{haw03} Hawkins, E., Maddox, S., \& Merrifield, M. 2002, \mnras, 336,
L13
\bibitem[Hewitt \& Burbidge(1993)]{hb93}  Hewitt, A. \& Burbidge, G. 1993, \apjs, 87,
451
\bibitem[Karlsson(1977)]{kar77}  Karlsson, K. G. 1977, \aap, 58,
237
\bibitem[Karlsson(1990)]{kar90}  Karlsson, K. G. 1990, \aap, 239,
50
\bibitem[Napier \& Burbidge(2003)]{bn03}  Napier, W. M. \& Burbidge, G.  2003,
\mnras, 342, 601
\bibitem[Richards et al.(2002)]{ric02}  Richards, G. T., et al.  2002, \aj, 123,
2945
\bibitem[Schneider et al.(2003)]{sch03}  Schneider, D. P., et al.  2003, \aj, 126,
2579
\bibitem[Schneider et al.(2005)]{sch05}  Schneider, D. P., et al.  2005, accepted
by AJ (astro-ph/0503679)
\bibitem[Strauss et al.(2002)]{str02}  Strauss, M. A., et al.  2002, \aj, 124,
1810

\end{thebibliography}
\end{document}